\documentclass[aps,prl,twocolumn]{revtex4}

\usepackage{mathrsfs}
\usepackage{amsmath, amssymb, amsfonts, bbm}
\usepackage{graphics,epsfig,color,subfigure}

\newcommand{\be}{\begin{equation}}
\newcommand{\ee}{\end{equation}}
\def\bea{\begin{align}}
\def\ena{\end{align}}

\def\nnb{\nonumber}
\def\Tr{\mbox{ Tr }}
\def\beqa{\begin{eqnarray}}
\def\enqa{\end{eqnarray}}
\def\nnb{\nonumber}
\def \Nc{N_c}
\def\One{\rm \mathbbm{1}}

\begin{document}

\title{Model Independent Tests of Skyrmions and Their Holographic Cousins}

\author{Aleksey Cherman}
\email{alekseyc@physics.umd.edu}
\affiliation{Maryland Center for Fundamental Physics, Department of Physics, University of Maryland,
College Park, MD 20742-4111}

\author{Thomas D. Cohen}
\email{cohen@physics.umd.edu}
\affiliation{Maryland Center for Fundamental Physics, Department of Physics, University of Maryland,
College Park, MD 20742-4111}

\author{Marina Nielsen}
\email{mnielsen@if.usp.br}
\affiliation{Instituto de F\'{\i}sica, Universidade de S\~{a}o Paulo,
C.P. 66318, 05389-970 S\~{a}o Paulo, SP, Brazil}

\begin{abstract}
We describe a new exact relation for large $N_c$ QCD for the long-distance behavior of baryon form factors in the chiral limit.  This model-independent relation is used to test the consistency of the structure of several baryon models.  All 4D semiclassical chiral soliton models satisfy the relation, as does the Pomarol-Wulzer holographic model of baryons as 5D Skyrmions.  However, remarkably, we find that the holographic model treating baryons as instantons in the Sakai-Sugimoto model does not satisfy the relation.
\end{abstract}

\maketitle

In the large $\Nc$ limit, QCD becomes a weakly-interacting theory of long-lived mesons, and baryons appear as soliton-like configurations of meson fields\cite{LargeN}.  There is a class of baryon models, the semiclassical chiral soliton models such as the Skyrme model\cite{Skyrme, ANW}, which use these large $\Nc$ properties of baryons to motivate their structure.  In these models, one constructs nontrivial hedgehog configurations of meson fields (``Skyrmions"), and identifies baryons with the quantum states of the collective motions of the Skyrmions.  While the models do not fully describe QCD, they encode fundamental features associated with the dynamics of QCD in the large $\Nc$ and chiral limits.  These features are reflected in certain  model-independent relations of large $\Nc$ QCD which all these models satisfy\cite{ANW, largeNrelations, CohenChiral,CohenIsovector}.

There have recently been proposals to use the gauge/gravity duality\cite{GaugeGravity} to construct holographic models of baryons.  In these models, one constructs a five-dimensional gravitational theory which is supposed to be dual to a 4D field theory which is hoped to approximate QCD.   These models are supposed to make sense when the dual field theory is in the large $\Nc$ limit, and is strongly coupled with large 't Hooft coupling $\lambda = g^2 \Nc$.  Some of the bulk fields are associated with meson fields of the dual field theory, and baryons are modeled as quantum states of topologically nontrivial configurations of these bulk fields.  The details of the new models look very different from traditional 4D chiral soliton models.
Thus, it is interesting to check whether these new holographic models of baryons capture large $\Nc$ and chiral physics correctly by checking various model-independent baryon relations.

In this letter a new model-independent relation for baryons is obtained.  It is sensitive to the anomalous coupling of the baryon current to three pions and becomes an exact result of QCD in the combined large $N_c$ and chiral limits (with the large $N_c$ limit taken first).   This relation is studied in Skyrme-like chiral soliton models and used to test two new holographic models of baryons\cite{PomarolWulzer,InstantonBaryons}.   It would be interesting to check the relation in other holographic models as well\cite{HologModels}.  The relation is for a ratio of position-space electric and magnetic baryon `form factors' , in the large distance limit $r \rightarrow \infty$:
\be
\label{Rrelation}
\lim_{r\rightarrow \infty} \frac{\tilde{G}_E^{I=0} \tilde{G}_E^{I=1}}{\tilde{G}_M^{I=0} \tilde{G}_M^{I=1}} = \frac{18}{r^2} \;,
\ee
where isoscalar and isovector electric and magnetic position-space form factors are defined as
\begin{align}
\label{defFormFactors}
\tilde{G}_E^{I=0}(r) &= \frac{1}{4\pi} \int~d\Omega\langle p\uparrow|J_{I=0}^{0}|p\uparrow\rangle  \\
\tilde{G}_M^{I=0}(r) &= \frac{1}{4\pi} \int~d\Omega{1\over2}\varepsilon_{i j 3}\langle p \uparrow|x_{i} J_{I=0}^{j}|p\uparrow\rangle \\
\tilde{G}_E^{I=1}(r) &= \frac{1}{4\pi} \int~d\Omega\langle p\uparrow|J_{I=1}^{\mu = 0, a = 3} | p \uparrow\rangle \\
\tilde{G}_M^{I=1}(r) &= \frac{1}{4\pi} \int~d\Omega{1\over2} \varepsilon_{i j 3}\langle p\uparrow| x_{i} J_{I=1}^{\mu=j, a=3} | p \uparrow\rangle
\end{align}
where $| p\uparrow\rangle$ represents a properly normalized wave function for
proton state of spin up, and  $J^{\mu}_{I=0}, J^{\mu a}_{I=1}$ are the isoscalar and isovector currents, where $a$ is an isospin index. These position-space form factors can be related to the standard experimentally accessible momentum-space form factors by
\begin{align}
\tilde{G}_E^{I=0,I=1}(r) &= \int{\frac{d^{3}q} {(2\pi)^3}~e^{i\vec{q} \cdot \vec{r}}G_E^{I=0,I=1}} (q)\nnb\\
\tilde{G}_M^{I=0,I=1}(r) &= \frac{-i} {3}\int{\frac{d^{3}q} {(2\pi)^3}~e^{i\vec{q} \cdot \vec{r}}~\vec{q}} \cdot  \vec{r}~G_M^{I=0,I=1} (q)\nnb
\end{align}
Unlike more traditional long-distance probes such as moments of form factors ({\it e.g.,} charge radii), the position space form factors are finite in the chiral limit.


{\it Skyrmions.} We begin by deriving Eq.~(\ref{Rrelation}) in the context of the
the original Skyrme model for massless pions,  which is a nonlinear sigma model in which the soliton is stabilized via the inclusion of a four derivative term. The model only contains pions; they are parameterized by the unitary matrix $U=\exp{(i\vec{\pi}.\hat{r}/f_\pi)}$. A static Skyrmion solution is obtained by minimizing the classical energy with the pions in a hedgehog ansatz: $\vec{\pi}=F(r)\hat{r}$. Baryons are identified with the quantum states of a slowly rotating Skyrmion, so that, for instance, the nucleon-delta mass splitting $\Delta$ is encoded in the moment of inertia of the Skyrmion.  The static properties of nucleons in the Skyrme model were first discussed in Ref.~\cite{ANW}.   Using the expressions for the angle-averaged isovector and isoscalar currents given in Ref.~\cite{ANW}, it is not hard to show that the position-space form factors as defined above are given by
\begin{align}
\label{anwff}
G_E^{I=0}(r) = -{\sin^2F~F' \over 2\pi^2r^2}, &\;\; G_M^{I=0}(r)= -{\sin^2F~F' \over 12\pi^2 \cal{I}}\\
G_E^{I=1}(r) = {\Lambda \over6 {\cal I }}, &\;\;  G_M^{I=1}(r) ={\Lambda \over18} \nnb
\end{align}
in the Skyrme model, where $\cal{I}$ is the moment of inertia of the Skyrmion and $\Lambda$ is a function which for small $F$  (corresponding to  long distances)  is $\Lambda=4 f_{\pi}^2 F^2 + {\cal O}(F^4)$.  At large distances $F(r) = \eta/r^2$, the pion field evaluated in a nucleon state behaves as a massless p-wave Yukawa potential; $\eta$ fixes the strength of the pion tail in the model.  Using the fact that in the Skyrme model $g_{A} = \frac{8\pi}{3} f_{\pi}^{2} \eta $, while $\Delta = \frac{3}{2\cal{I}}$, we see that the long-distance behavior of the form factors is
\begin{align}
\label{FF_GSE}
\lim_{r\rightarrow \infty} G_{I=0}^{E} &= \frac{3^3}{2^9\pi^5} \frac{1}{f_{\pi}^3}\left(\frac{g_A}{f_{\pi}}\right)^3 \frac{1}{r^9}\\
\label{FF_GSM}
\lim_{r\rightarrow \infty} G_{I=0}^{M} &= \frac{3\Delta}{2^9  \pi^5}\frac{1}{f_{\pi}^{3}} \left(\frac{g_A}{f_{\pi}}\right)^3 \frac{1}{r^7}\\
\label{FF_GVM}
\lim_{r\rightarrow \infty} G_{I=1}^{M} &= \frac{1}{2^5 \pi^2} \left(\frac{g_A}{f_{\pi}}\right)^2 \frac{1}{r^4}\\
\label{FF_GVE}
\lim_{r\rightarrow \infty} G_{I=1}^{E} &= \frac{\Delta}{2^4  \pi^2} \left(\frac{g_A}{f_{\pi}}\right)^2 \frac{1}{r^4} \;.
\end{align}
Taking the appropriate ratio then yields Eq.~(\ref{Rrelation}).


{\it Model-independence.} It is trivial to see that the previous result will hold for {\it any} semiclassical chiral soliton model that correctly incorporates the anomalous coupling of the baryon current to three pions. While the details of the form factors will differ, the large distance behavior will be given  by Eqs.~(\ref{FF_GSE})-(\ref{FF_GVE}) in all models, since these equations depend only on the structure of the currents at long distance, and the fact that $F(r) = \eta/r^2$ at long distance. 

There is considerable experience with relations which hold for all chiral soliton models treated semiclassically.  In all known cases\cite{largeNrelations,CohenChiral,CohenIsovector} these are in fact generic relations of large $N_c$ QCD and are fully independent of model-dependent assumptions.  For observables sensitive to the chiral dynamics of pions a subtlety arises: the large $N_c$ and chiral limits do not generally commute. The semiclassical treatment of the soliton amounts to taking the large $N_c$ limit first\cite{CohenChiral,CohenIsovector}.  Thus, one fully expects Eq.~(\ref{Rrelation}) to hold for QCD in the the large $N_c$  limit, with the chiral limit taken after $\Nc \rightarrow \infty$.

The model independence of these QCD relations was shown by large $N_c$ $\chi PT$.  It is known that the long distance hadronic physics of QCD is dominated by the pion cloud, and unlike in the meson sector, for baryons pion loops contribute at leading order in the $1/N_c$ expansion\cite{CohenLeinweber} so that large $\Nc$ baryon physics can be described by heavy baryon $\chi PT$.  At the longest distance it depends only on the low energy constants $g_A$, $f_\pi$ and $m_\pi$.  At large $N_c$, the Delta nucleon is  parametrically low lying,  and thus $\Delta \equiv M_{\Delta} -M_N \sim N_c^{-1}$ also serves as a low energy constant.   The longest range observables can thus be expressed entirely in terms of $g_A$, $f_\pi$, $\Delta$ and $m_\pi$ in a manner that is fixed by large $N_c$ $\chi PT$.

\begin{figure}
        \label{ChiPT}
        \centering
        \subfigure[Isoscalar]{
            \label{ChiPT:scalar}
            \includegraphics[width=1.5in]{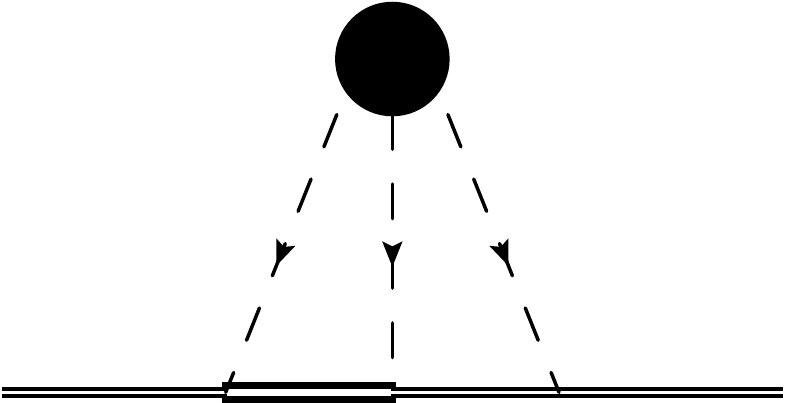}
        }
        \subfigure[Isovector]{
            \label{ChiPT:vector}
            \includegraphics[width=1.5in]{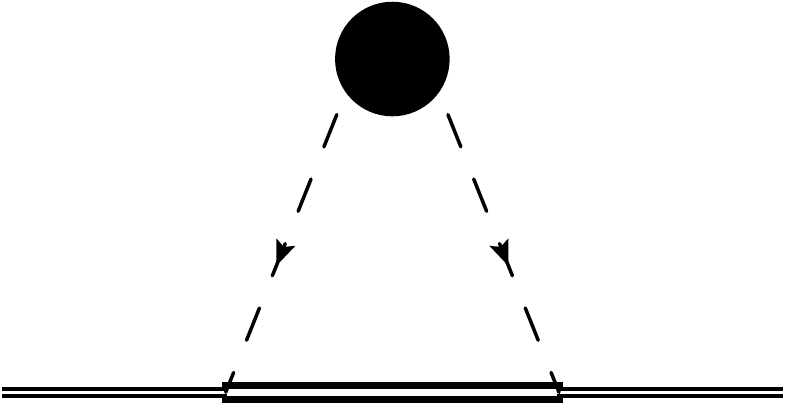}
        }
    \caption{Typical diagrams contributing to the long-distance behavior of form factors in large $\Nc$ $\chi PT$.  In Fig.~\ref{ChiPT:scalar}, an EM current probes the nucleon isoscalar form factors through a three-pion interaction from the anomaly.  In Fig.~\ref{ChiPT:vector} an EM current probes the nucleon isovector form factors through a two-pion interaction.  Intermediate states can be either nucleons or deltas, which are shown as double lines above. }
\end{figure}

For the present case,  note that the leading contributions to the isoscalar electric and magnetic form-factors are both proportional to $f_{\pi}^{-3} (g_{A}/f_{\pi})^3$ from the pion couplings and the anomaly structure (see  Fig.~\ref{ChiPT:scalar}), while the leading contribution to the isovector electric and magnetic form factors are both proportional to $(g_{A}/f_{\pi})^2$ from the pion couplings (see Fig.~\ref{ChiPT:vector}).  The electric isovector and magnetic isoscalar form factors are $1/N_c$ suppressed compared to the magnetic isovector  and isoscalar and are each proportional to $\Delta$\cite{CohenIsovector}.  The ratio in Eq.~(\ref{Rrelation}) was chosen to have all of the dependence on $g_{A}, f_{\pi}$ and $\Delta$ cancel, so that the result is simply a universal number times a factor of $1/r^2$, independent of any low energy constants; thus it is model independent.

{\it 5D Skyrmions. } We now study the behavior of form factors in the Pomarol-Wulzer model\cite{PomarolWulzer,PanicoWulzer}, a holographic model of baryons.  In the spirit of bottom-up holographic models of QCD\cite{AdS_QCD}, the gravity dual background has the line element $ds^2 = a(z)^2  \left( dx^{\mu} dx_{\nu} - dz^2\right)$, where $a(z) = L/z$, $z \in [0,L]$.
The model includes two 5D U(2) gauge fields, $\bold{L}_{M}$ and $\bold{R}_{M}$, which are associated with the left and right chiral currents $\bar{q}_{R,L} \gamma_{\mu} t_{a} q_{R,L}$ in the dual 4D theory via the AdS/CFT dictionary\cite{GaugeGravity}.  The QCD-like  4D theory lives on the UV boundary at $z=0$ of the dual gravity theory.   The action of the model simply consists of the gauge kinetic terms\footnote{We have set the constant $\alpha$ of Ref.~\cite{PomarolWulzer,PanicoWulzer} to unity, as it must be in large $\Nc$ QCD, although Eq.~(\ref{Rrelation}) will be satisfied even with $\alpha \neq 1$\cite{CCN2}.}, as well as a 5D Chern-Simons term to incorporate the effects of the chiral anomaly in the QCD-like dual field theory.  The action of the 5D model is given by
\begin{align}
\label{s5d}
S &= -\frac{M_5}{2} \int{d^{5}x \sqrt{g} \Tr[ \bold{L}^2_{MN}  + \bold{R}^2_{MN}]  }   \\
&+{-i N_c\over24\pi^2} \int_{5D}{\left[\omega_5(\bold{L})-\omega_5(\bold{R})\right]}\nnb,
\end{align}
where $\bold{L}_{MN}, \bold{R}_{MN}$ are the U(2) gauge field strengths, $M,N = z,\mu $, $\omega_5$ is the Chern-Simons 5-form, and $M_5 \sim {\sc O}(\Nc^1)$ is an input parameter of the model.

In this model baryons are identified as the quantum states of slowly rotating `5D Skyrmions'.   The 5D Skyrmions are defined to be topologically nontrivial configurations of the 5D gauge fields with baryon number $B = 1$, where
\be
B = \frac{1}{32\pi^2} \int{d^{3}x \, dz \, \epsilon_{\hat{\mu} \hat{\nu} \hat{\rho} \hat{\sigma} }\Tr( \bold{L}^{\hat{\mu} \hat{\nu}} \bold{L}^{\hat{\rho} \hat{\sigma}} - \bold{R}^{\hat{\mu} \hat{\nu}} \bold{R}^{\hat{\rho} \hat{\sigma}} )  }
\ee
and $\hat{\mu}=x_i,  z$.  The hedgehog-like field configurations with $B=1$ can be parametrized in terms of five functions: $\phi_{1}(r,z)$, $\phi_{2}(r,z)$, $A_{1}(r,z)$, $A_{2}(r,z)$, $s(r,z)$.  If we split the U(2) gauge fields into $SU(2)$ and $U(1)$ parts as $\bold{R}_M = R^{a}_{M} \sigma^{a}/2 + \hat{R}_M \One/2$ (and similarly for $\bold{L}_M$), the ansatz that defines the functions above can be written as $L^{a}_{i}(\bold{x},z) = - R^{a}_{i}(-\bold{x},z)$, $L^{a}_{5}(\bold{x},z) = R^{a}_{i}(\bold{x},z)$, $\hat{L}_{0}(\bold{x},z) = \hat{R}^{a}_{0}(-\bold{x},z)$,
with all other components set to zero, and
\begin{align}
R^{a}_{j} &= - \frac{1+\phi_{2} (r,z)}{r^2} \epsilon_{j a k} x_k + \frac{\phi_{1}(r,z)}{r^3} (r^2 \delta_{j a} - x_j x_a) \nnb \\
 & \; \; + \frac{A_{1}(r,z)}{r^2} x_j x_a; \;
R^{a}_{5} =  \frac{A_{2}(r,z)}{r^2} x_a, \; \; \hat{R}_{0} = \frac{s(r,z)}{r}. \nnb 
\end{align}
The CS term stabilizes the 5D Skyrmion size to ${\cal O}(L)$.  

Baryons are identified with the quantum states of slowly rotating 5D Skyrmions, as described in detail in Ref.~\cite{PanicoWulzer}.   To check if the model satisfies Eq.~(\ref{Rrelation}), we need the isoscalar and isovector currents of the model, which can be determined from the left and right chiral currents.  The left chiral currents are $J^{a}_{L \mu} = M_5 (a(z) L^{a}_{\mu 5})_{z=0}$, $\hat{J}_{L\mu} = M_5 (a(z) \hat{L}^{a}_{\mu 5})_{z=0}$,
and the right chiral currents are defined similarly. By writing the isospin currents for the explicit case of a slowly rotating 5D Skyrmion with B=1, it is possible to show that
\begin{align}
G_E^{I=0}(r) &= -{4\over N_c} M_5 \left[ a(z)\partial_z s\over r \right]_{z=0}, \nnb\\
G_M^{I=0}(r)&= -{2\over 3N_c\cal{I}} M_5 \left[ r a(z)\partial_z Q~ \right]_{z=0}, \nnb\\
G_E^{I=1}(r) &= {2\over 3\cal{I}}M_5 \left[a(z)(\partial_z v-2 (\partial_z\chi_2- A_2 \chi_1)) \right]_{z=0}, \nnb\\
G_M^{I=1}(r) &= -{4\over9}M_5 \left[ a(z)(\partial_z \phi_2 - A_2 \phi_1)\right]_{z=0},
\label{cur5D}
\end{align}
where $\cal{I}$  is the moment of inertia, and $v(r,z)$, $Q(r,z)$, $\chi_{1}(r,z)$, $\chi_{2}(r,z)$ parametrize the collective rotations of the 5D Skyrmion, and are defined in Ref.~\cite{PanicoWulzer}.

To check Eq.~(\ref{Rrelation}), we need to know the large $r$ behavior of the form factors above.  The large $r$ and small $z$ asymptotics of the 5D Skyrmion solutions were determined in Ref.~\cite{PanicoWulzer}.  Using these asymptotic solutions, it is easy to see that
\begin{align}
\label{5Dff}
\lim_{r\rightarrow \infty} G_{I=0}^{E}  &= -{\beta^3L^6\over\pi^2}{1\over r^9}, \; \; \lim_{r\rightarrow \infty} G_{I=0}^{M} = {\beta^3L^6\over6\pi^2\cal{I}}{1\over r^7},\\
\lim_{r\rightarrow \infty} G_{I=1}^{E}  &= {8\beta^2\over3\cal{I}}M_5L^3{1\over r^4},\;\;  \lim_{r\rightarrow \infty} G_{I=0}^{M} = -{8\beta^2\over9}M_5L^3{1\over r^4} \nnb,
\end{align}
where $\beta$ is a parameter that can be determined numerically from the equations of motion of the 5D Skyrmion. Finally, using Eq.~(\ref{5Dff}) it is clear that Eq.~(\ref{Rrelation}) is satisfied in the Pomarol-Wulzer model of baryons as 5D Skyrmions.  It is also not hard to see that  Eq.~(\ref{FF_GSE})-(\ref{FF_GVE}) are satisfied in this model. This means that the Pomarol-Wulzer model correctly captures the large $\Nc$ chiral physics of QCD to which Eq.~(\ref{Rrelation}) and  Eq.~(\ref{FF_GSE})-(\ref{FF_GVE}) are sensitive.


{\it Baryons as holographic instantons.} Refs.~\cite{InstantonBaryons, HRYY, HSS,Zahed} propose a holographic model associating baryons with instantons in a 5D holographic theory arising from the Sakai-Sugimoto model\cite{SS}, and calculate the electromagnetic form factors of the nucleon.    The action of the model is obtained from the leading terms of a derivative expansion of the D-brane effective action and is 
\be
S =-\kappa\int{d^{4}x dz\Tr \frac{\bold{F}_{\mu\nu}^2}{2 k(z)^{1/3} }+ k(z) {\bold{F}}_{\mu z}^2  +{N_c\over24\pi^2}\int\omega_5({\bold{A}})} \nnb
\ee
where $k(z) = 1+z^2$, $\bold{A}_M$ is a 5D $U(N_f)$ gauge field, $\bold{F}_{MN}$ is the field
strength, $\kappa = {\lambda N_c / (216\pi^3)}$, $\lambda = g_{YM}^2 \Nc$ is the  't Hooft coupling,
and $\omega_5({\bold{A}})$ is the Chern-Simons 5-form.   The semiclassical treatment of the model is justified in the large $\Nc$ and large $\lambda$ limits.  The model has a single mass scale $M_{KK}$, which has been set to unity.

In this model, baryons are identified with the quantum states of a slowly rotating instanton with
instanton number $1$ on the four-dimensional space parametrized
by $x^{\hat{\mu}},~\hat{\mu}=1,2,3,z$.  The Chern-Simons term stabilizes the instanton size to be of order $\lambda^{-1/2}$.  Unfortunately, as noticed in Ref.~\cite{InstantonBaryons}, this means that as  $\lambda \rightarrow \infty$ the theory becomes sensitive to higher-derivative terms that are not included in the action.  Such concerns make it important to test whether the model as implemented is consistent with known large $\Nc$ chiral physics.

To check whether this model satisfies Eq.~(\ref{Rrelation}), we use the formalism of Ref.~\cite{HSS} and express the position-space form factors in terms of the fields and parameters of the Sakai-Sugimoto model.  In Eqs.~(2.96)-(2.99) of Ref.~\cite{HSS}, Hashimoto, Sakai, and Sugimoto
derive the expressions for the scalar and isovector currents. Using the
expressions for the currents, which are derived using the fact that the instantons are localized arbitrarily well at $z=0$ in the large $\lambda$ limit, and the definitions in
Eq.~(\ref{defFormFactors}) we get:
\begin{align}
\label{SS_form_factors}
G_E^{I=0}(r) &=-\sum_{n=1}^\infty g_{v^n} \psi_{2n-1}(0) Y_{2n-1}(r),\\
G_M^{I=0}(r) &= - {9 \pi r \over 4 \lambda \Nc  }\sum_{n=1}^\infty g_{v^n} \psi_{2n-1}(0)  m_{2n-1}Y_{2n-1}(r) \nnb\\
G_E^{I=1}(r) &=-\sum_{n=1}^\infty g_{v^n}\psi_{2n-1}(0) Y_{2n-1}(r),\nnb\\
G_M^{I=1}(r) &=-\frac{\Nc}{3} \sqrt{\frac{2}{15}} \sum_{n=1}^\infty g_{v^n}\psi_{2n-1} (0) \rho_{2n-1}Y_{2n-1}(r) \nnb,
\end{align}
where $\left\{\psi_n(z)\right\}$ is a complete set
of functions normalized so that $\psi(z)\sim \kappa^{-1/2}$ that satisfy $-k(z)^{1/3} \partial_z (k(z)\partial_z \psi_n (z) ) = \rho^2_n \psi_n (z)$, where the eigenvalues $\rho^{2}_{n}$ (with $\rho_{n+1} >\rho_n$) are related to the masses of the vector mesons in this model by $m^{2}_{n} = \rho^{2}_{n} M_{KK}^{2}$, the vector meson decay constants $g_{v^n} = 2\kappa\lim_{z\to\infty} z \psi_{2n-1}(z)$, and $Y_n(r)$ are Yukawa
potentials $Y_{n}(r)=-e^{- \rho_n r}/(4\pi r)$.

Taking the large $r$ limit of the form factors in Eq.~(\ref{SS_form_factors}) we find that
\begin{align}
\label{SS_large_r}
\lim_{r\rightarrow \infty} G_E^{I=0}(r) &= \frac{g_{v^1} \psi_{1}(0) }{4\pi r}  e^{- \rho_1 r}, \\
\lim_{r\rightarrow \infty} G_M^{I=0}(r) &= {9 \pi r \over 16 \pi \lambda \Nc  } g_{v^1}\psi_1(0) \rho_1 {e^{-\rho_1 r}},\nnb\\
\lim_{r\rightarrow \infty} G_E^{I=1}(r) &= \frac{g_{v^1} \psi_{1}(0) }{4\pi r}  e^{- \rho_1 r}, \nnb \\
\lim_{r\rightarrow \infty} G_M^{I=1}(r) &= \frac{\Nc}{12\pi} \sqrt{\frac{2}{15}}   g_{v^1}\psi_1(0){\rho_1}~{e^{-{\rho_1}r}}\nnb  \;,
\end{align}
which makes it easy to see that
\be
\lim_{r \rightarrow \infty} \frac{\tilde{G}_E^{I=0} \tilde{G}_E^{I=1}}{\tilde{G}_M^{I=0} \tilde{G}_M^{I=1}} = \frac{\lambda \sqrt{40/3}  }{\pi \rho^{2}_{1} r^2} \; .
\ee

The ratio is sensitive to model parameters, so Eq.~(\ref{Rrelation}) is not obeyed and the model does not correctly encode the large $\Nc$ chiral properties of baryons.  This is troubling in that the ability to describe chiral symmetry and its spontaneous breaking are supposed to be principal virtues of the model. In fact, one can see a symptom of the problem arising already in Eq.~(\ref{SS_form_factors}).  The form factors depend only on couplings to vector mesons, and not to pions, in contradiction to large $\Nc$ $\chi PT$.  This makes the failure of the model to satisfy Eq.~(\ref{Rrelation})  unsurprising.    While the symptoms of the problem are clear, whether they represent a technical difficulty in the implementation of the model or a deeper structural problem remains an open question.  It appears likely that the issue is connected to a non-commutativity of the large $\lambda$ and chiral limits in this model, as will be discussed in an upcoming work\cite{CCN2}.  

{\it Acknowledgements.} A.C. and T.D.C. acknowledge the support of the US Dept. of
Energy, the hospitality of the INT during the workshop ``New frontiers in Large N gauge theories", and thank D.~T.~Son, A.~Wulzer, K.~Hashimoto, T.~Sakai, and S.~Sugimoto for illuminating conversations.  M.N. acknowledges the support of CNPq and FAPESP and the hospitality and support of the Theoretical Hadronic Physics
Group during her stay at the University of Maryland.

\end{document}